\documentclass[twocolumn,epsfig]{revtex4}
\usepackage{epsfig,color}
\begin{document}
\title{Observable to explore high density behaviour of symmetry energy}

\author{Aman D. Sood$^1$ }
\email{amandsood@gmail.com}
\address{
$^1$SUBATECH,
Laboratoire de Physique Subatomique et des
Technologies Associ\'ees \\University of Nantes - IN2P3/CNRS - Ecole des Mines
de Nantes 
4 rue Alfred Kastler, F-44072 Nantes, Cedex 03, France}
\date{\today}

\maketitle

\section*{Introduction}
The nuclear matter equation of state (EOS) of asymmetric nuclear matter has attracted a lot of
attention recently \cite{li08}. The EOS of asymmetric nuclear matter can be described
approximately by
\begin{equation}
E(\rho, \delta) = E_{0}(\rho,
\delta=0)+E_{\textrm{sym}}(\rho)\delta^{2}
\end{equation}
where
 $\delta$ = $\frac{\rho_{n}-\rho_{p}}{\rho_{n}+\rho_{p}}$ is isospin asymmetry,
  E$_{0}$($\rho$, $\delta$) is the energy of pure symmetric nuclear matter, and E$_{\textrm{sym}}$($\rho$)
  is the symmetry energy. The symmetry energy is important not only to the nuclear physics community as it sheds light on the structure of radioactive nuclei, reaction dynamics induced
by rare isotopes, but also to astrophysicists since it acts as a
probe for understanding the evolution of massive stars and the
supernova explosions. Although $E_{sym}$ at saturation density is known to be around 30 MeV its value at higher densities is poorly known. Experimentally, symmetry energy is not a
directly measurable quantity and has to be extracted from
observables which are related to symmetry energy. Therefore, a
crucial task is to find such observables which can shed light on
symmetry energy. In this paper, we aim to see the
sensitivity of collective transverse in-plane flow to symmetry
energy at low as well as high densities and also to see the effect of different density
dependencies of symmetry energy on the same. The various forms of
symmetry energy used in present study are: E$_{\textrm{sym}}
\propto (u)$, E$_{\textrm{sym}} \propto (u)^{0.4}$, and
E$_{\textrm{sym}} \propto (u)^{2}$, where \emph{u} =
$\frac{\rho}{\rho_{0}}$. The different density
dependencies of symmetry energy are shown in Fig. 1.
The present study is carried out using IQMD model which is described briefly in the next section.

\section*{The model}
In IQMD \cite{hart98} model the hadrons propagate using Hamilton equations of motion: 

\begin {eqnarray}
\frac{d\vec{{r_{i}}}}{dt} = \frac{d\langle H
\rangle}{d\vec{p_{i}}};& & \frac{d\vec{p_{i}}}{dt} = -
\frac{d\langle H \rangle}{d\vec{r_{i}}}~where
\end {eqnarray}
\begin {eqnarray}
\langle H\rangle& =&\langle T\rangle+\langle V \rangle
\nonumber\\
& =& \sum_{i}\frac{p^{2}_{i}}{2m_{i}} + \sum_{i}\sum_{j>i}\int
f_{i}(\vec{r},\vec{p},t)V^{ij}(\vec{r}~',\vec{r})
 \nonumber\\
& & \times f_{j}(\vec{r}~',\vec{p}~',t) d\vec{r}~ d\vec{r}~'~
d\vec{p}~ d\vec{p}~'.
\end {eqnarray}
 The baryon potential\emph{ V$^{ij}$}, in the above relation, reads as
 \begin {eqnarray}
  \nonumber V^{ij}(\vec{r}~'-\vec{r})& =&V^{ij}_{Skyrme} + V^{ij}_{Yukawa} + V^{ij}_{Coul} 
  \nonumber\\
   & & + V^{ij}_{mdi} + V^{ij}_{sym}.
 \end {eqnarray}
 Where the potential terms in eq. 4 represent, respectively, Skyrme, Yukawa, Coulomb, momentum dependent interaction, and symmetry potential. 
\begin{figure}[!t] \centering
\vskip 0.5cm
\includegraphics[angle=0,width=6cm]{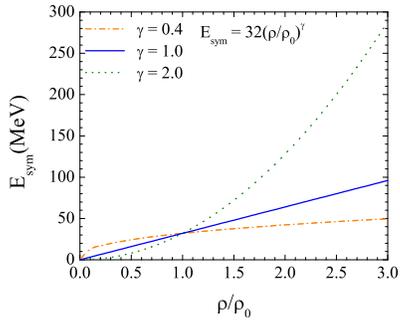}
\caption{\label{fig1} Different forms of symmetry energy.}
\end{figure}

\section*{Results and Discussion}
There are several methods used in the literature to define the
nuclear transverse in-plane flow. In most of the studies, one uses
($p_{x}/A$) plots where one plots ($p_{x}/A$) as a function of
$Y_{\textrm{c.m.}}/Y_{\textrm{beam}}$. Using a linear fit to the
slope, one can define the so-called reduced flow (F).
Alternatively, one can also use a more integrated quantity
``directed transverse in-plane flow
$\langle{p_{x}^{\textrm{dir}}}\rangle$'' which is defined as \cite{gautam7}:
\begin{equation}
\langle p_{x}^{\textrm{dir}}\rangle~=~\frac{1}{A}\sum_i {\rm
sign}\{Y(i)\}~p_{x}(i),
\end{equation}
where $Y(i)$ and $p_{x}(i)$ are the rapidity distribution and
transverse momentum of the $ith$ particle. In this definition, all
rapidity bins are taken into account. It therefore presents an
easier way of measuring the in-plane flow than complicated
functions such as ($p_{x}/A$) plots.
\par
 In Fig. 2 we display $<\frac{p_{x}}{A}>$ as a function of
Y$_{\textrm{c.m.}}/Y_{\textrm{beam}}$ at final time (left panels)
and the time evolution of $<p_{x}^{\textrm{dir}}>$ (right panels)
calculated at 100 (top panel), 400 (middle) and 800 MeV/nucleon
(bottom) for different density dependencies of symmetry energy.
Solid, dash dotted, and dotted lines represent the symmetry energy
proportional to $\rho$, $\rho^{0.4}$ and $\rho^{2}$, whereas
dashed line represents calculations without symmetry energy.
Comparing the left and right panels in Fig. 2, we find that both
the methods show similar behavior to symmetry energy. For example,
at incident energy of 100 MeV/nucleon for E$_{\textrm{sym}}
\propto \rho^{0.4}$, $<p_{x}^{\textrm{dir}}>$ = 0. Similarly, the
slope of $<\frac{p_{x}}{A}>$ at midrapidity is zero. We also find
that the transverse momentum is sensitive to symmetry energy and
to its density dependences F$_{1}(u)$, F$_{2}(u)$ and F$_{3}(u)$
in the low energy region (100 MeV/nucleon) only. At energies above
Fermi energy, both the methods show insensitivity to the different
symmetry energies. 
\begin{figure}[!t] \centering
\vskip 0.5cm
\includegraphics[angle=0,width=6cm]{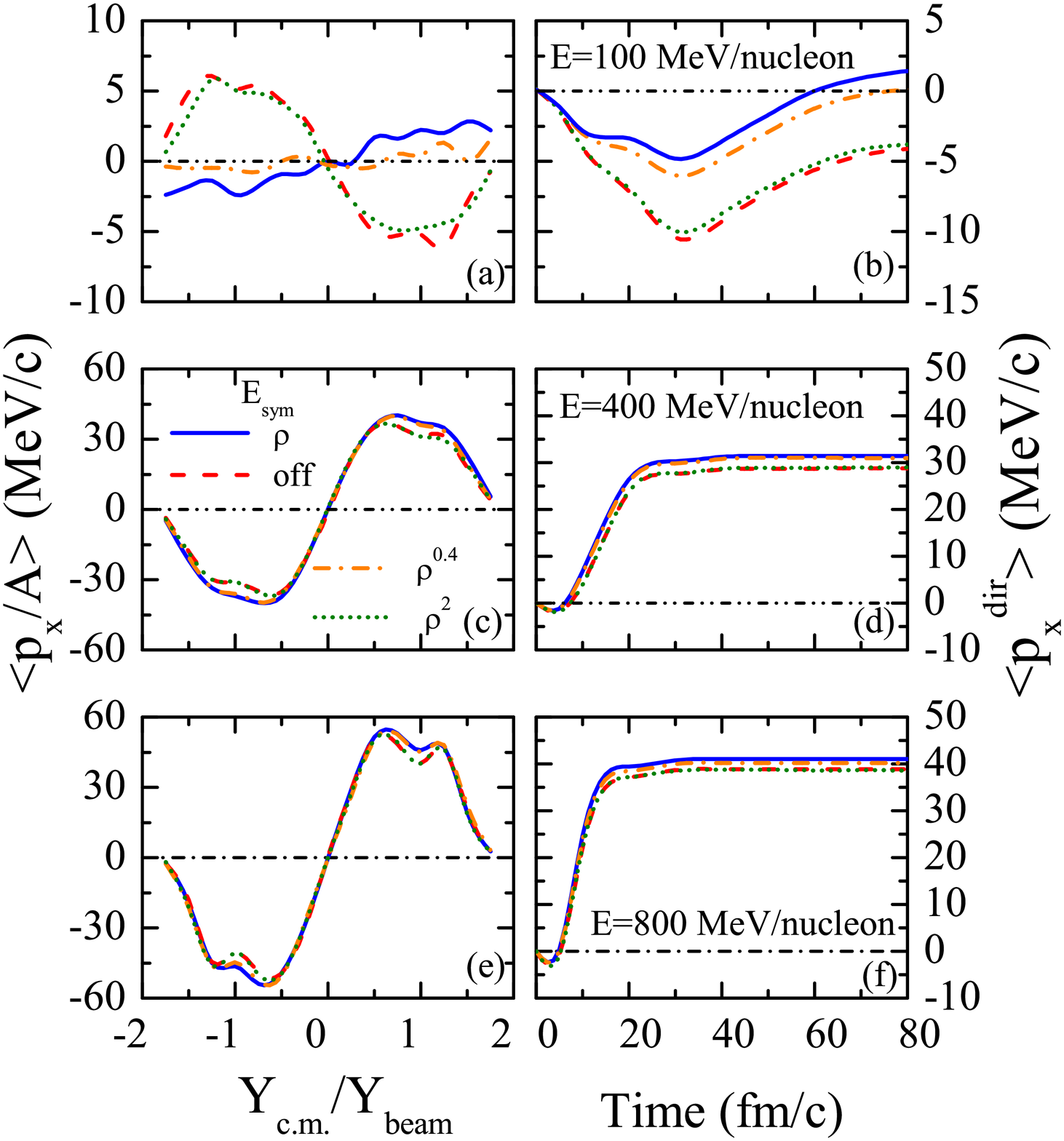}
\caption{\label{fig1} Left panel:
$<\frac{p_{x}}{A}>$ as a function of
Y$_{\textrm{c.m.}}/Y_{\textrm{beam}}$ and $<p_{x}^{\textrm{dir}}>$ at different incident energies. Panels and lines are explained in the
text.}
\vskip -0.5cm
\end{figure}
This could be because (i) the repulsive nn
scattering dominates the mean field at high energies. (ii) As explained in ref \cite{gautam7} that the effect of $E_{sym}$ on nucleons during the initial stages of the reaction (0-30 fm/c) decide the fate of final state $<p_{x}^{\textrm{dir}}>$. At low energies duration of high dense phase will prevail for a longer duration alllowing the $E_{sym}$ to affect the flow. Whereas at high energies, although maximum density reached will be larger but the duration of high density phase will be very small not providing  $E_{sym}$ with sufficient time to make an impact on flow.
\section*{Acknowledgments}
This work has been supported by a grant from Indo-French Centre for the Promotion of Advanced Research (IFCPAR) under project no 4104-1.



\end{document}